# On Defining Smart Cities using Transformer Neural Networks

Andrei Khurshudov

Yale University, School of Engineering and Applied Science, New Haven, USA

address:  andrei.khurshudov@yale.edu

**Abstract**

Cities worldwide are rapidly adopting "smart" technologies, transforming urban life. Despite this trend, a universally accepted definition of "smart city" remains elusive. Past efforts to define it haven't yielded a consensus, as evidenced by the numerous definitions in use. In this paper, we endeavored to create a new "compromise" definition that should resonate with most experts previously involved in defining this concept and aimed to validate one of the existing definitions. We reviewed 60 definitions of smart cities from industry, academia, and various relevant organizations, employing transformer architecture-based generative AI and semantic text analysis to reach this compromise. We proposed a semantic similarity measure as an evaluation technique, which could generally be used to compare different smart city definitions, assessing their uniqueness or resemblance. Our methodology employed generative AI to analyze various existing definitions of smart cities, generating a list of potential new composite definitions. Each of these new definitions was then tested against the pre-existing individual definitions we've gathered, using cosine similarity as our metric. This process identified smart city definitions with the highest average cosine similarity, semantically positioning them as the closest on average to all the 60 individual definitions selected.

Introduction

Let's begin with the following existing definition: a smart city is a technologically advanced urban area that uses information and communication technology (ICT) and the Internet of Things (IoT) to enhance its sustainability and efficiency, as well as to improve the quality of its services. Its fundamental mission is to enhance the quality of life, economic competitiveness, and sustainability [1,2]. We will refer to this as "Definition 0.1".

Smart cities employ electronic methods and sensors to collect data, which is then utilized to manage assets, resources, and services efficiently. This data, gathered from citizens, devices, buildings, and assets, is processed, and analyzed to improve city operations. This optimization extends to traffic and transportation systems, power plants, utilities, water supply networks, waste management, crime detection, and the functioning of public services such as schools, libraries, and hospitals. At the heart of the smart city concept is the integration of ICT and physical devices connected to the IoT network. This integration aims to enhance the efficiency of city operations and foster better connectivity with citizens.

One of the foundational papers that significantly influenced the smart city concept was "City of Bits: Space, Place, and the Infobahn," written by Mitchell in 1995 [3]. This seminal work delved into the integration of digital technology and information networks within urban spaces, essentially laying the groundwork for what would later be known as smart cities. While the specific term "smart city" may not have been explicitly used in these early discussions, the foundational ideas and principles concerning urban technology and connectivity were clearly articulated.

In his 2008 article, "Will the Real Smart City Please Stand Up?", Hollands [4] critically examined the frequent claims by cities to be "smart." He observed that many cities label themselves as such due to the incorporation of Information and Communication Technologies (ICTs) in their operations, often without a clear definition or evidence to substantiate these claims. This trend, he noted, arises globally, with little deep consideration of what truly constitutes a smart city.

Even today, despite extensive discussions and a wealth of publications, a universally accepted definition of "smart cities" remains elusive, leading to a wide array of interpretations. The following two examples illustrate the variability and differing focuses prevalent in these definitions:

- A smart city is an urban area that uses an array of digital technologies to enrich residents' lives, improve infrastructure, modernize government services, enhance accessibility, drive sustainability, and accelerate economic development. Smart cities are the cities of the future [5].
- A smart city is a utopian vision of a city that produces wealth, sustainability, and well-being by using technologies to tackle wicked problems [6].

The lack of a universally agreed-upon definition of a "smart city" carries several practical implications, which are significant for urban planning, policy-making, and technological implementation. These include:

1. Public Misunderstanding: Misconceptions about smart cities can lead to public dissatisfaction and lack of support.
2. Varying Standards and Expectations: Differing views on what constitutes a smart city can lead to variable standards and key performance indicators (KPIs), complicating the comparison and assessment of smart city initiatives.
3. Inconsistent Implementation: Varied interpretations result in uneven implementation of smart city solutions, affecting quality of life improvements and technological progress.
4. Difficulty in Policy Formulation: The absence of a clear framework challenges policymakers in devising effective smart city policies and regulations.
5. Challenges in Securing Funding: Unclear definitions hinder city planners in securing funding, as investors may be unsure of project outcomes and benefits.
6. Technology Misalignment: Solutions from technology providers may not fully align with city needs, leading to inefficiencies and underutilization.
7. Collaboration Challenges: Absence of a unified definition impedes collaboration and best practice sharing among cities, providers, and stakeholders.

In response to the challenge of converging on a singular "best" definition of a smart city, studies have been conducted, resulting in the proposal of various new definitions.

For example, the ITU's Focus Group on Smart Sustainable Cities conducted a thorough analysis, as detailed by Al-Nasrawi et al. [7]. After examining approximately 120 definitions, they proposed the following new definition: "A smart sustainable city is an innovative city that uses information and communication technologies (ICTs) and other means to improve quality of life, efficiency of urban operations and services, and competitiveness, while also ensuring that it meets the needs of present and future generations in terms of economic, social, and environmental aspects."

Ali and Panchal [8] conducted an analysis of over 100 definitions. At the conclusion of their study, they proposed a new, "futuristic" definition: "A smart city is defined by its ability to learn from experience (E) in relation to a specific task (T), a performance measure (P), and resource optimization (O). If its performance in task (T), as measured by (P), improves in terms of resource optimization (O) through experience (E), then the city is deemed smart."

Gracias et al. [9] analyzed 36 different definitions. They observed that smart cities integrate various domains such as transportation, energy, health, education, and governance, fostering an interconnected and intelligent urban ecosystem. Utilizing word frequency analysis, they proposed a new definition: "Smart cities employ digital and communication technologies along with data analytics to create an efficient, effective service environment that enhances urban quality of life and promotes sustainability."

Despite the plethora of existing definitions and multiple efforts to consolidate them, none have achieved universal acceptance. This might be attributed to the divergent viewpoints of different stakeholders involved in smart city development, and perhaps their reluctance to reach a compromise.

In response, the primary objective of this paper was to propose a definition of a smart city that can become widely accepted. We have utilized technologies such as generative AI, large language models (LLM), and transformer-based semantic similarity analysis to identify or to compose the definition that most closely aligns with the extensive range of existing definitions, thereby best encapsulating the concept of smart cities. This approach enhances the likelihood of widespread acceptance of this definition.

**Methodology**

In this study, we aimed to, as much as possible, eliminate human bias by relying solely on AI and Generative AI techniques such as text summarization and semantic similarity analysis. For these, we utilized the GPT-4 Large Language Model (LLM) [10], a leading transformer-based generative AI solution now that was released in 2023. GPT, which stands for Generative Pretrained Transformer, refers to a type of advanced neural network designed for natural language processing. Originating from the paper "Attention Is All You Need" by Vaswani et al. [11], transformers are distinguished by their unique architecture, especially the "attention" mechanism. This mechanism enables the LLM to assess the importance of different parts of input data, such as focusing on specific words or phrases in a sentence that are crucial for understanding the context or for the task at hand. This capability allows transformers to "comprehend" and generate contextually relevant language, making them exceptionally useful for semantic (contextual) analysis.

The term "pretrained" in GPT indicates that it has undergone initial training on a vast corpus of text, allowing it to understand a broad spectrum of language patterns and structures. This extensive training equips GPT with exceptional proficiency in tasks involving human communication, such as answering questions, translating languages, and generating coherent, contextually appropriate text. This capability optimizes it for various applications that demand sophisticated, human-like language processing abilities.

Initially, we conducted a comprehensive literature review to compile a diverse list of 60 definitions of a smart city (see Appendix). These definitions were sourced from various entities, including companies, organizations, and academic papers. The complete list of these definitions can be found in the Appendix.

Secondly, we conducted a semantic similarity analysis [12] between our baseline Definition 0.1 and the 60 individual definitions compiled. For this purpose, we utilized cosine similarity analysis, a method commonly employed in semantic analysis to measure the similarity between two text documents or vectors. This technique is based on the cosine of the angle between two vectors in a multi-dimensional space, representing the respective text documents. The cosine similarity metric calculates the cosine of the angle between these vectors, yielding a value ranging from -1 to 1. A value nearing 1 indicates a high degree of similarity, suggesting that the documents share numerous common words or similar word distribution patterns. On the other hand, a value close to -1 or 0 suggests low similarity. A value of 1 indicates identical sentences, 0 indicates no similarity, and -1 indicates complete dissimilarity. This approach is invaluable in natural language processing for various tasks, including document clustering, information retrieval, and plagiarism detection, as it effectively discerns the likeness in content between different texts based on word usage patterns.

The cosine similarity between two sentences, or text vectors, is calculated using the following formula:

$$\text{Cosine Similarity}(A, B) = \frac{\sum_{i=1}^{n} A_i \times B_i}{\sqrt{\sum_{i=1}^{n} A_i^2} \times \sqrt{\sum_{i=1}^{n} B_i^2}} \quad (1)$$

Where:

- A and B are two sentences being compared.
- $A_i$ and $B_i$ are the components (e.g., term frequency weights) of sentences A and B in the vector space, respectively.
- n is the number of dimensions in the vector space. In the context of sentences, this usually corresponds to the number of unique terms across both sentences after some preprocessing like tokenization.

The numerator of the equation is the dot product of the two sentence vectors. The denominator is the product of the Euclidean norms (or lengths) of these vectors. The resulting cosine similarity value (score) quantifies the degree of similarity between the sentences. This measure effectively captures the linguistic and semantic similarities or differences between two text segments.

For our analysis, we utilized "Sentence Transformers," a cutting-edge transformer-based Python framework for generating sentence and text embeddings [13] using pre-trained models [14]. This framework can produce semantically meaningful sentence embeddings, which are highly useful in unsupervised learning scenarios such as semantic textual similarity analysis, clustering, and semantic search. The foundational methodology of "Sentence Transformers" was detailed in the paper by Reimers and Gurevych [15]. In our research, we analyzed these embeddings using the cosine similarity approach. This approach allows for a nuanced measurement of the semantic similarity between sentences (or smart city definitions).

We utilized a specific pre-trained model from the "Sentence Transformers" framework, known as "all-mpnet-base-v2" [16]. This top-ranked model is designed as an all-rounder, optimized for a wide range of use cases. It has been trained on an extensive and diverse dataset comprising over 1 billion training pairs. This extensive training enables the model to generate high-quality embeddings, albeit with a trade-off in processing speed.

Third, we employed the GPT-4 engine, to summarize all 60 individual definitions of smart cities. Utilizing GPT's powerful text summarization capabilities, we generated 20 different composite definitions derived from the semantic content of these individual 60 definitions. Generative AI solutions like GPT-4 excel at text summarization.

Within the realm of Large Language Models (LLMs) like GPT-4, there are two primary types of text summarization approaches – "extractive" and "abstractive":

- Extractive Summarization: This method involves selecting exact key phrases or sentences from the original text to form a summary. It maintains the original phrasing and context.

- Abstractive Summarization: This method rephrases and condenses the original text, creating a new, shorter version that captures the essence of the information.

Each summarization method is selected based on the specific requirements of the task, including factors like accuracy, coherence, and the characteristics of the text to be summarized. While extractive summarization, if applied to all 60 individual definitions, would retain the exact phrases from each source, this approach could result in an overly lengthy, cumbersome, and impractical compilation. Therefore, we opted for abstractive summarization. This method allowed us to synthesize the essence of each definition into a more concise, coherent, and practical format, effectively capturing the underlying concepts without the constraints of the original texts" structure.

The "temperature" setting in large language models, including GPT, is a crucial parameter influencing the randomness or creativity of the generated text. Higher temperatures lead to more varied and creative outputs, while lower temperatures yield more predictable and conventional responses. In our study, the GPT model's temperature was set to 0.7. This level of temperature allowed for a balanced output, providing enough randomness to generate diverse responses during text summarization, yet maintaining a level of predictability to ensure the responses remain relevant and coherent.

An interaction with Large Language Models like GPT-4 was carried out using textual requests or prompts. For the abstractive summarization process, we have provided all 60 initial definitions and employed the following prompt: "Produce one complete definition of a smart city using various definitions mentioned below. Only include technology-related characteristics mentioned in these definitions. Write in one paragraph, shorten as much as possible. Limit to 35 words." This prompt was designed to guide the AI in generating a concise, technology-focused definition of a smart city, synthesizing key elements from a range of existing definitions while adhering to a word limit for brevity and focus.

Here's an overview of how GPT-4 summarized 60 individual definitions into one composite definition:

1. First, GPT-4 processes the input text, which, in this case, comprises the 60 individual definitions.

2. Then, the model tokenized the text and computes the probability of each token in the corpus of occurring next, Prob(word | context) (probability of a word in context). It utilizes temperature and top probabilities to select the next token. Then it repeats this process again and again by adding the last predicted token to the context. It continues until it reaches a stopping token like "." or newline "/n".

3.  Using a technique known as the "attention mechanism," GPT-4 identified the parts of the input that are most relevant to the overarching theme. It then generated a summary that includes these key elements, ensuring that the summary is coherent, concise, and captures the essence of the input.

This method was used to generate the first composite (summary) definition out of the 20 composite definitions. The remaining 19 were generated using an additional prompt: "Produce another 19 different definitions, limit each to 35 words." For each of these definitions, the GPT-4 model referenced all 60 individual definitions. However, it randomized the word (token) completion process during each iteration, resulting in 20 different, albeit thematically consistent, answers. This approach ensured that each generated definition was unique, while still being grounded in the semantic content of the 60 individual definitions.

Forth, we conducted another round of semantic similarity comparison. This time, it was between the 20 newly generated composite definitions and the original set of 60 individual definitions. We generated a series of embeddings for 20 new composite definitions (see below) and the 60 individual definitions and conducted a cosine similarity comparison between each pair of these embeddings, utilizing Python code for this computation. This step allowed us to assess how closely each composite definition aligned with the broader range of existing individual definitions.

Finally, we undertook additional analyses, comparisons, and rule extraction. We compared the composite definitions to each other to measure their commonalities.

**Results and Discussion**

*Generating composite definitions*

Below are all 20 composite definitions that were generated from the 60 individual ones (see Appendix).

1.  A smart city effectively uses digital and sensor-based technologies to collect comprehensive data about its people and surroundings, applying this information to elevate urban living standards, enhance operational efficiency, and foster sustainable and economic growth.
2.  In a smart city, advanced ICT and IoT systems gather and interpret data about urban life and infrastructure, driving improvements in city management, service quality, and resident well-being, while addressing environmental and economic challenges.
3.  A smart city is an urban landscape where technology, particularly through sensors and data analytics, plays a key role in optimizing resource management, improving public services, and ensuring sustainable development.
4.  Such a city integrates cutting-edge digital solutions and interconnected networks to manage its operations, enhance public services, and ensure the welfare and quality of life of its citizens, balancing technological advancements with sustainable urban living.
5.  A smart city harnesses ICT and IoT for data-driven insights, leading to more efficient city operations, better quality of citizen services, sustainable practices, and responsive governance to urban challenges.
6.  This city model utilizes a network of sensors and ICT infrastructure to analyze urban dynamics, aiming to improve efficiency in utilities, transportation, and public services while fostering an inclusive and sustainable environment.
7.  A smart city combines technology, data, and innovation to enhance its infrastructure and services, focusing on sustainability, economic development, and a higher quality of life, adapting continuously to emerging urban challenges.
8.  Employing a blend of digital technologies, smart cities process and utilize data from their environments and inhabitants to streamline operations, improve services, and create a resilient, sustainable, and livable urban future.
9.  In a smart city, the synergy of sensors, data analytics, and ICT solutions transforms urban areas into more efficient, sustainable, and livable spaces, with a focus on enhancing the overall urban experience and addressing key challenges.

10. A smart city strategically employs technology, particularly ICT and IoT, to gather, analyze, and utilize data for optimizing urban functions, enhancing living conditions, and driving sustainable and economic advancements.
11. It's an urban area that leverages digital innovation and sensor technology to improve infrastructure management, citizen engagement, and resource efficiency, contributing to a sustainable and economically vibrant community.
12. A smart city is defined by its use of interconnected technologies and data analysis to enhance public services, infrastructure efficiency, and quality of life, while proactively addressing environmental and social challenges.
13. Such a city integrates digital sensors and ICT to monitor and improve urban systems, facilitating better decision-making for city governance, sustainable development, and improved citizen welfare.
14. A smart city uses a network of digital technologies and data collection methods to make urban areas more efficient, livable, and responsive to the needs of its residents and the environment.
15. It's an urban ecosystem where technology-driven solutions are applied to optimize city operations, enhance public services, and ensure sustainable urban development.
16. A smart city combines digital infrastructure, data analytics, and ICT to manage urban resources and services more effectively, aiming to create a responsive, sustainable, and high-quality urban life.
17. In a smart city, the integration of technology and data is key to improving urban efficiency, environmental sustainability, and the overall quality of life for its inhabitants.
18. It's a city model where ICT and sensor-based data collection are utilized to streamline city operations, improve citizen services, and achieve economic and environmental sustainability.
19. A smart city is an urban area where digital and sensor technologies are integrated into the city fabric, enabling smarter resource management, efficient public services, and a higher standard of living for its residents.
20. A smart city harnesses advanced technologies, including ICT and IoT, to collect and analyze data from its infrastructure and residents. This approach is key to enhancing urban efficiency, improving quality of life, fostering sustainable development, and driving economic growth, making the city more responsive and adaptable to the needs of its inhabitants and the environment.

Each of these 20 composite definitions appears to encapsulate some aspects of the message conveyed in the individual 60.

*Similarity analysis. Definition 0.1*

Initially, we compared the original Definition 0.1 with 59 individual definitions to calculate the average cosine similarity value. This definition itself, being part of the initial 60, was excluded from the list for this specific comparison. The analysis resulted in an average similarity score of 0.854, suggesting a strong similarity of Definition 0.1 with the considered list. It is worth noting that when it was not excluded, the similarity score increased insignificantly to 0.856. This suggests that each individual definition in the list of 60 had a proportionally small effect on the overall result. Subsequently, we extended our analysis to include the composite 20 definitions. Also, notice that – without the second sentence, the definition's similarity score against 59 declines slightly to 0.848.

*Similarity analysis. 20 new composite definitions.*

The ideal definition for our purpose is the one that demonstrates, on average, the highest similarity to all (60) pre-existing individual definitions.

Figure 1 displays the similarity scores obtained from comparing each of the 20 composite definitions against each of the 60 individual definitions. The range of similarity score values is quite broad, extending from as low as 0.38, indicating poor similarity, to as high as 0.97, signifying very high similarity. This variation highlights the diverse degrees of alignment between the composite and individual definitions.

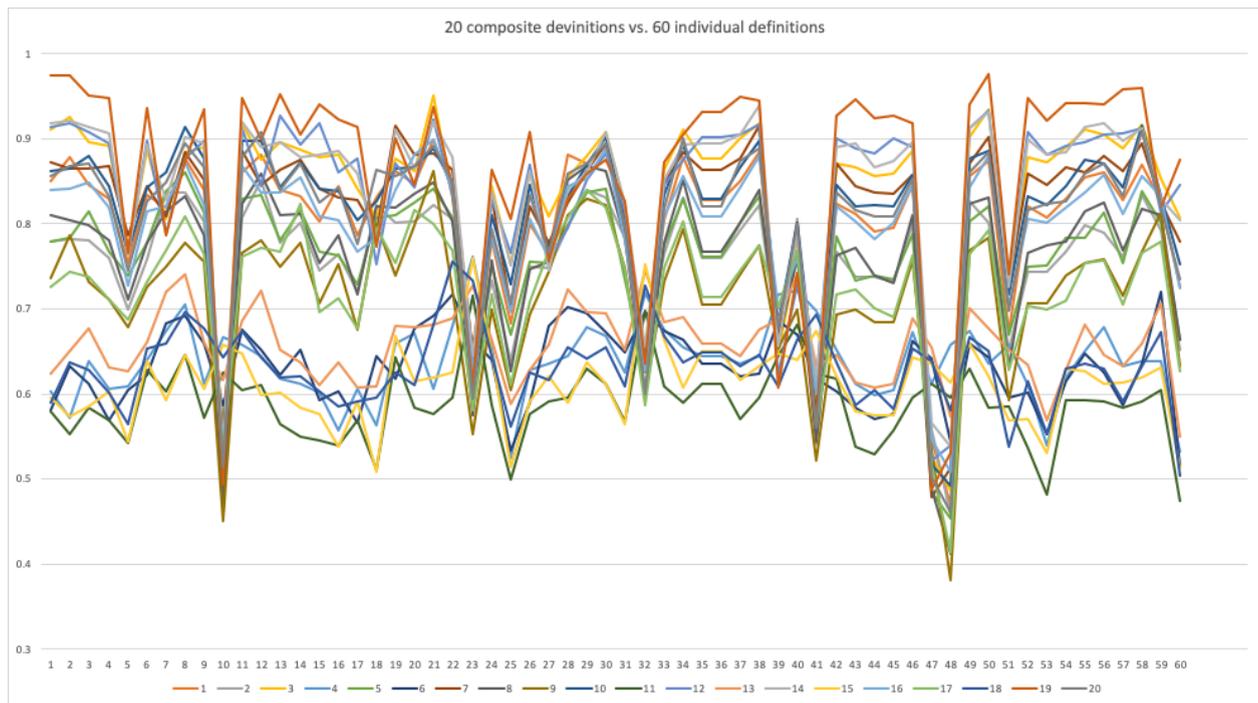

Figure 1. Similarity scores obtained from comparing each of the 20 composite definitions against each of the 60 individual definitions.

It's worth noting that certain composite definitions consistently scored low in terms of similarity metrics, such as definitions 11, 4, 6, 15, and 18, while others frequently ranked at the top of the chart, notably definitions 19, 14, 3 and 12.

We also observed that some existing individual definitions, such as definitions 48, 10, and 47, exhibited significant variance from most of the initial definitions. This suggests that these particular definitions differed notably from the others created by their respective authors.

For reference, consider the case of composite definition 9 and individual definition 48, which yielded the lowest similarity score in our analysis:

1. Composite definition 9: In a smart city, the synergy of sensors, data analytics, and ICT solutions transforms urban areas into more efficient, sustainable, and livable spaces, with a focus on enhancing the overall urban experience and addressing key challenges.

2. Individual definition 48: A city that invests in human and social capital, political participation of citizens, management of natural resources, and traditional and modern networked infrastructure [17].

The similarity score calculated for these two definitions was 0.38. This low score indicates that the two definitions likely address very different priorities or aspects of the smart city concept, reflecting the wide range of interpretations and focuses on the field.

For comparison, the following is an example of high semantic similarity (score of 0.98) between composite definition 19 and initial definition 50. One can observe that these two definitions appear to focus on similar subjects or themes:

1. Composite definition 19: A smart city is an urban area where digital and sensor technologies are integrated into the city fabric, enabling smarter resource management, efficient public services, and a higher standard of living for its residents.

2. Individual definition 50: A smart city is a system integration of technological infrastructure that relies on advanced data processing with the goals of making city governance more efficient, citizens happier, businesses more prosperous and the environment more sustainable [18].

Figure 2 illustrates the average performance of each composite definition, with the top four highlighted in different colors for distinction.

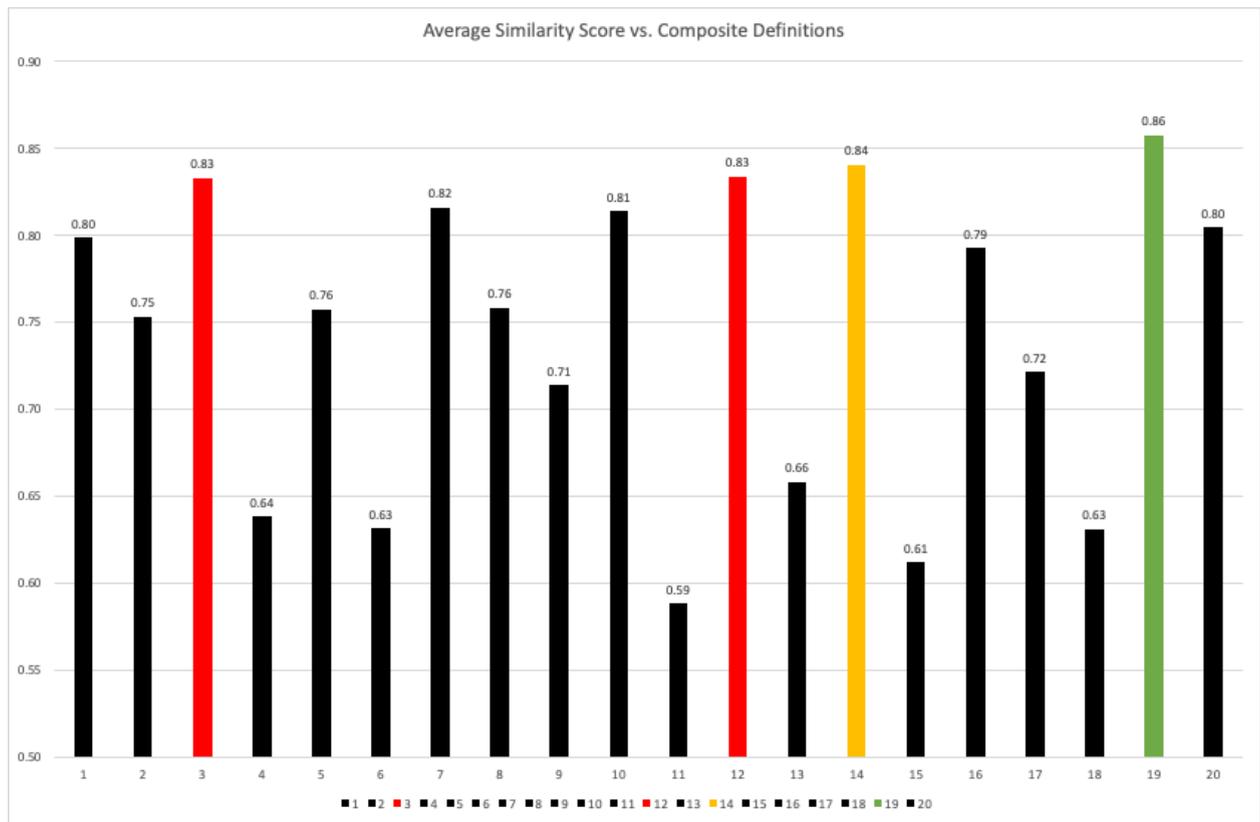

Figure 2. Average performance of each composite definition, with the top four highlighted in different colors for distinction.

Composite definition 19 emerges as the top performer among the 20 definitions, achieving an average cosine similarity score of 0.86. This score indicates that it is the most similar to the initial set of 60 definitions on average. Composite definitions 12 and 14, along with definition 3, are close contenders, with scores of 0.84 and 0.83, respectively. However, to achieve a concise top-3 ranking, we randomly selected definition 12 over 3, thereby reducing the list to three. Notably, all top-ranked definitions commonly focus on themes such as sensors," "digital," "data," "technologies," "efficiency," and so forth:

1. (19) A smart city is an urban area where digital and sensor technologies are integrated into the city fabric, enabling smarter resource management, efficient public services, and a higher standard of living for its residents.
2. (14) A smart city uses a network of digital technologies and data collection methods to make urban areas more efficient, livable, and responsive to the needs of its residents and the environment.
3. (12) A smart city is defined by its use of interconnected technologies and data analysis to enhance public services, infrastructure efficiency, and quality of life, while proactively addressing environmental and social challenges.

Table 1. Semantic similarity scores across the top-3 composite definitions.

| Definition | 19 | 14 | 12 |
|---|---|---|---|
| 19 | 1 | 0. 936 | 0.932 |
| 14 | 0.936 | 1 | 0.93 |
| 12 | 0. 932 | 0. 93 | 1 |

Table 1 provides a direct comparison of the semantic similarity scores for the top-3 composite definitions. This comparison confirms their contextual closeness, suggesting that any one of them could be selected as "the definitive smart city definition" for universal use.

*Similarity analysis. Definition 0.1 vs. 19.*

Our analysis indicates that definitions 19 and 0.1 have nearly identical average semantic similarities with the 60 (or 59) individual definitions we collected, scoring 0.858 and 0.854, respectively.

To further confirm their semantic similarity, we directly compared both definitions and found a high similarity score of 0.96, affirming their closeness.

Finally, we evaluated the three definitions previously suggested in the introduction for semantic similarity against the 60 individual definitions, as detailed in Table 2.

Table 2: indicates that definitions 19 and 0.1, with average similarity scores of 0.85 – 0.86, more effectively represent the average meaning contained in the 60 individual smart city definitions we selected.

| Definition source | Average Cosine similarity score | Comments |
| --- | --- | --- |
| Al-Nasrawi et al [7]. | 0.80 | |
| Ali and Panchal [8]. | 0.79 | Removed letters in parenthesis |
| Gracias at all [9]. | 0.78 | |
| Definition 0.1 | 0.854 | |
| Definition 19 | 0.858 | |

As a result, Definition 0.1 can be considered validated, and Definition 19 can be proposed as its equally valid but more concise alternative.

*Proposed future steps.*

As we acknowledge that the "smart city" concept will continue to evolve with the introduction of new technologies, services, and objectives, it is imperative to establish a formal process for determining how its definition will evolve accordingly.

The method outlined above can be applied in the future to evaluate any new smart city definitions against the corpus of pre-existing definitions (see Appendix and definitions 19, 14, and 12). This proposal is akin to the "moving average" approach in statistics, where the new estimate incorporates past history.

Such approach ensures the gradual evolution of the definition, thereby helping to avoid abrupt shifts or discontinuities.

**Conclusions**

With numerous smart definitions available, ambiguity in the field can hinder agreement on key performance indicators (KPIs) and measurement of a city's "smartness," ultimately impeding smart city efforts. The objective of this study was to leverage transformer architecture-based generative AI and semantic similarity text analysis to converge on the most acceptable definition of a smart city. The main conclusions are as follows:

1. The initial definition of a smart city (Definition 0.1) performs commendably when compared to 20 artificially generated composite definitions. It demonstrates a high average semantic similarity score of 0.854 with the list of 59 initial definitions, comparable to the score of the best composite definition, 19 (0.858), and can be considered validated. Thus, Definition 0.1 is recommended for use:

    "A smart city is a technologically advanced urban area that uses information and communication technology (ICT) and the Internet of Things (IoT) to enhance its sustainability and efficiency, as well as to improve the quality of its services. Its fundamental mission is to enhance the quality of life, economic competitiveness, and sustainability."

2. While the above definition is effective, it might be perceived as lengthy and challenging to memorize. By contrast, Definition 19, which has a similarity score of 0.858 with 60 definitions and a very high semantic similarity of 0.96 with Definition 0.1, is notably more concise, consisting of only 34 words as opposed to 52. Therefore, we argue that Definition 19 is superior to any other definition considered in this study, making it our top recommendation for defining a smart city:

   "A smart city is an urban area where digital and sensor technologies are integrated into the city fabric, enabling smarter resource management, efficient public services, and a higher standard of living for its residents."

3. Given the probability of new definitions emerging in the future, we propose employing the aforementioned technique to evaluate these new contenders against both the list of known definitions (Appendix) and the best composite definitions identified in this study (19, or even 14 and 12). Such approach ensures the gradual evolution of the definition, thereby helping to avoid abrupt shifts or discontinuities.

**Appendix: A list of 60 initial existing smart city definitions.**

1. A smart city is an urban area that uses an array of digital technologies to enrich residents' lives, improve infrastructure, modernize government services, enhance accessibility, drive sustainability, and accelerate economic development. Smart cities are the cities of the future [5].

2. A smart city is a city in which a suite of sensors typically hundreds or thousands) is deployed to collect electronic data from and about people and infrastructure so as to improve efficiency and quality of life [19].

3. A smart city is a municipality that uses information and communication technologies (ICT) to increase operational efficiency, share information with the public and improve both the quality of government services and citizen welfare [20].

4. A smart city is a city that uses technology to provide services and solve city problems. A smart city does things like improve transportation and accessibility, improve social services, promote sustainability, and give its citizens a voice [21].

5. The main goals of a smart city are to improve policy efficiency, reduce waste and inconvenience, improve social and economic quality, and maximize social inclusion [21].

6. A smart city is one that uses technology to efficiently engage citizens and meet their needs [22].

7. Smart cities use information and communication technologies (ICT) to improve the ways they operate [23].

8. A smart city uses information and communication technology (ICT) to improve operational efficiency, share information with the public and provide a better quality of government service and citizen welfare. The main goal of a smart city is to optimize city functions and promote economic growth while also improving the quality of life for citizens by using smart technologies and data analysis. The value lies in how this technology is used rather than simply how much technology is available [24].

9. A smart city is a concept that sees the adoption of data-sharing smart technologies including the Internet of Things (IOT) and information communication technologies (ICTs) to improve energy efficiency, minimize greenhouse gas emissions, and improve quality of life of a city's citizens [25].

10. A broad, integrated approach to improving the efficiency of city operations, the quality of life for its citizens, and growing the local economy [26].

11. A smart city is a framework, predominantly composed of Information and Communication Technologies (ICT), to develop, deploy, and promote sustainable development practices to address growing urbanization challenges [27].

12. A smart city uses digital technology to connect, protect, and enhance the lives of citizens. IoT sensors, video cameras, social media, and other inputs act as a nervous system, providing the city operator and citizens with constant feedback so they can make informed decisions [28].

13. A smart city is defined as the effective integration of physical, digital and human systems in the built environment to deliver a sustainable, prosperous and inclusive future for its citizens [29].

14. A smart city is characterized by the integration of smart data, transportation, energy, infrastructure, and IoT (Internet of Things) [30].

15. A smart city is defined as the one that adopts scalable solutions that take advantage of ICT to increase efficiencies, reduce costs, and enhance quality of life [31].

16. A smart city is typically a kind of municipal Internet of Things — a network of cameras and sensors that can see, hear and even smell. These sensors, especially video cameras, generate massive amounts of data that can serve many civic purposes like helping traffic flow smoothly [32].

17. A Smart City as an urban area that has become more efficient and/or more environmentally friendly and/or more socially inclusive through the use of digital technologies. The goal of a Smart City is to improve its attractiveness to citizens and/or businesses by enhancing and/or adding city services [33].

18. The smart city, like the smart home, is built on and around the "Internet of things," in which networked products gather, store, and share user data while communicating with one another in order to create improved and highly-efficient living environments [34].

19. A Smart city uses technology and innovation to improve the urban environment – leading to improved quality of life, greater prosperity and sustainability, and engaged and empowered citizens [35].

20. A smart city connects the physical infrastructure, the information-technology infrastructure, the social infrastructure, and the business infrastructure to leverage the collective intelligence of the city [36].

21. A Smart City is one in which sensor-driven data collection and powerful analytics are used to automate and orchestrate a wide range of services in the interests of better performance, lower costs and lessened environmental impact [37].

22. A smart city model uses modern information and communication technology to improve its infrastructure, share information with the public and provide a better city living experience [38].

23. City that increases the pace at which it provides social, economic and environmental sustainability outcomes and responds to challenges such as climate change, rapid population growth, and political and economic instability by fundamentally improving how it engages society, applies collaborative leadership methods, works across disciplines and city systems, and uses data information and modern technologies to deliver better services and quality of life to those in the city (residents, businesses, visitors), now and for the foreseeable future, without unfair disadvantage of others or degradation of the natural environment [39].

24. A smart sustainable city is an innovative city that uses information and communication technologies (ICTs) and other means to improve quality of life, efficiency of urban operation and services, and competitiveness, while ensuring that it meets the needs of present and future generations with respect to economic, social and environmental aspects [7].

25. IDC: smart city is one that embraces technology for urban transformation to meet social, financial and environmental outcomes; a definition that takes an outcomes-based view of modernization and digitization in states, provinces, counties, cities and towns [40].

26. Smart city is the one that employs ICT to fulfill market demand, i.e., the citizens.; An ultra-modern urban area that addresses the needs of businesses, institutions, and especially citizens [41].

27. Smart and sustainable cities are expected to form a cornerstone for achieving resource efficiency and sustainability worldwide [42].

28. Smart city uses sensor technology and intelligent technologies to realize automatic, real-time operations, and comprehensive perception of urban operations on the basis of Digital City [43].

29. Smart cities are comprised of diverse and interconnected components constantly exchanging data and facilitating improved living for a nation's population [44].

30. A smart city employs a combination of data collection, processing, and disseminating technologies in conjunction with networking and computing technologies and data security and privacy measures encouraging application innovation to promote the overall quality of life for its citizens and covering dimensions that include: utilities, health, transportation, entertainment, and government services [45].

31. Smart cities are: 1) sensible (sensors sense the environment) 2) connectable (networked devices bring the sensed information to the Web) 3) accessible (information on our environment is published and is accessible by users on the Web) 4) ubiquitous (users can access information at any time and in any place, while moving) 5) sociable (users acquiring information can publish it through their social network) 6) shareable (sharing is not limited to data, but also to physical objects that may be used when they are in free status), and 7) visible/augmented (the physical environment is retrofitted and information is seen not only by individuals through mobile devices, but also in physical places such as street signs [46].

32. Cities that contain intelligent things which can intelligently automatically and collaboratively enhance life quality, save people's lives, and act as sustainable resource ecosystems [47].

33. A smart sustainable city is an innovative city that uses information and communication technologies (ICTs) and other means to improve quality of life, the efficiency of urban operations and services, and competitiveness, while ensuring that it meets the needs of the present and future generations concerning economic, social and environmental aspects [48].

34. Smart cities are aimed to efficiently manage growing urbanization, energy consumption, maintain a green environment, improve economic and living standards of their citizens and raise people's capabilities to efficiently use modern information and communication technology (ICT) [49].

35. Smart cities employ information and communication technologies to improve: the quality of life for its citizens, local economy, transport, traffic management and interaction with government [50].

36. A smart city is a system that enhances human and social capital wisely using and interacting with natural and economic resources via technology-based solutions and innovations to address public issues and efficiently achieve sustainable development and high quality of life [51].

37. Smart city is an urban environment that utilizes ICT and other related technologies to enhance performance efficiency of regular city operations and quality of services (QoS) provided to urban citizens [52].

38. A smart city utilizes urban informatics and technologies for providing city services on a larger scale. It offers improved quality of life and a variety of innovative services such as energy, transport, healthcare, etc. [53].

39. Connecting the physical, IT, social, and business infrastructures to leverage the collective intelligence of the city [54].

40. In smart city architecture, information and communication technologies are used to improve living standards and its management by citizens and government [55].

41. The city that makes optimal use of all the interconnected information available today to better understand and control its operations and optimize the use of limited resources [56].

42. Smart City is Use/Innovation of Technology/ICT coupled with favorable government policies that promote the development of infrastructure, ease of doing business and citizen engagement leading to sustainable economic growth and citizen satisfaction through improved quality of life [57].

43. Smart Cities are those that integrate information communications technology across three or more functional areas. More simply put, a Smart City is one that combines traditional infrastructure (roads, buildings, and so on) with technology to enrich the lives of its citizens [58].

44. Smart city is the idea of creating a sustainable living environment along with state-of-the-art technology (ICT) integration."; "A smart city is a self-containing city that focuses on people's QoL above everything else [59].

45. A smart city has been generally defined as a developed urban area that uses information and technology (ICT), human capital and social capital in order to promote sustainable socio-economic growth and a high quality of life [60].

46. A smart city is a complex cyber-socio-technical system where human, cyber artifacts, and technical systems interact together to the purpose of achieving a goal related to the quality of life in urban areas [61].

47. A set of instruments across many scales that are connected through multiple networks and provide continuous data regarding people and environment in support of decisions about the physical and social form of the city [62].

48. A city that invests in human and social capital, political participation of citizens, management of natural resources, and traditional and modern networked infrastructure [17].

49. Smart cities are cities that balance economic, environmental, and societal advances to improve the wellbeing of residents through a widespread introduction of ICT and other technological tools [63].

50. A smart city is a system integration of technological infrastructure that relies on advanced data processing with the goals of making city governance more efficient, citizens happier, businesses more prosperous and the environment more sustainable. [18].

51. A city is smart if investments in human and social capital and traditional (transport) and modern (ICT) communication infrastructure fuel sustainable economic growth and a high quality of life, with a wise management of natural resources, through participatory governance [64].

52. A smart city is the concept could be briefly described as cities that "use information and communication technologies in order to increase the quality of life of their inhabitants while contributing to a sustainable development [65].

53. Smart city is "a futuristic approach to alleviate obstacles triggered by ever-increasing population and fast urbanization which is going to benefit the governments as well as the masses." Smart cities are "an endeavor to make cities more efficient, sustainable and livable [66].

54. A smart city is a utopian vision of a city that produces wealth, sustainability, and well-being by using technologies to tackle wicked problems [6].

55. Smart cities are urban areas in which information and communication technologies are used to solve their specific problems and support their sustainable development in social, economic and/or environmental terms [67].

56. A smart city is a place where traditional networks and services are made more efficient with the use of digital solutions for the benefit of its inhabitants and business [68].

57. A smart city is an urban area that uses technological or non-technological services or products that: enhance the social and ethical wellbeing of its citizens; provide quality, performance and interactivity of urban services to reduce costs and resource consumption; and increase contact between citizens and government [69].

58. A smart city is a technologically advanced urban area that uses information and communication technology (ICT) and the Internet of Things (IoT) to enhance its sustainability and efficiency, as well as to improve the quality of its services. Its fundamental mission is to enhance the quality of life, economic competitiveness, and sustainability [1].

59. Smarter Cities are urban areas that exploit operational data, such as that arising from traffic congestion, power consumption statistics, and public safety events, to optimize the operation of city services [36].

60. A smart city as one that has an active plan and projects in at least five of the eight functional areas of Energy, Buildings, Mobility, Technology, Infrastructure, Healthcare, Governance, and Citizens. Each of these key parameters has specific components that define the "Smartness" of a City [70].

*International Journal of Computer and Technology* Vol 24 (2024) ISSN: 2277-3061      https://rajpub.com/index.php/ijct